\documentclass[journal,10pt]{IEEEtran}

\ifCLASSINFOpdf
\else
  \usepackage[dvips]{graphicx}
\fi
\usepackage{url}
\usepackage[ruled,vlined]{algorithm2e}
\RestyleAlgo{ruled}
\usepackage{amsmath}
\usepackage{dsfont, cuted}
\usepackage{amsfonts}
\usepackage{amsthm}
\usepackage[utf8]{inputenc}
\usepackage{breqn}
\usepackage{xcolor}
\usepackage{subfigure}
\usepackage{multirow}
\usepackage{multicol}
\usepackage{booktabs}
\usepackage{fullwidth}
\usepackage[noadjust]{cite}
\usepackage{setspace}
\usepackage{hyperref}
\usepackage{bmpsize}
\usepackage{flushend}
\usepackage{graphicx}
\usepackage{pbalance}
\allowdisplaybreaks

\newtheorem{pro}{Proposition}
\newtheorem{thm}{Theorem}

\newtheorem{rem}{Remark}
\newtheorem{prf}{Proof}

\begin{document}

\title{RKFNet: A Novel Neural Network Aided Robust Kalman Filter}
\author{Pengcheng~Hao, Oktay~Karakuş,~\IEEEmembership{Member,~IEEE,} Alin~Achim,~\IEEEmembership{Senior Member,~IEEE}
    \thanks{This work was supported by the Chinese Scholarship Council (CSC).}
    \thanks{Pengcheng Hao and Alin Achim are with the Visual Information Laboratory, University of Bristol, Bristol BS1 5DD, U.K. (e-mail: ju18422@bristol.ac.uk; alin.achim@bristol.ac.uk)}
    \thanks{Oktay Karakuş is with the School of Computer Science and Informatics, Cardiff University, Abacws, Cardiff, CF24 4AG, UK. (e-mail: karakuso@cardiff.ac.uk)}
}

\markboth{IEEE Transactions on XXXX, Vol. xx, No. x, 2024}
{Shell \MakeLowercase{\textit{et al.}}: Bare Demo of IEEEtran.cls for IEEE Journals}
\maketitle

\begin{abstract}
Driven by the filtering challenges in linear systems disturbed by non-Gaussian heavy-tailed noise, the robust Kalman filters (RKFs) leveraging diverse heavy-tailed distributions have been introduced. However, the RKFs rely on precise noise models, and large model errors can degrade their filtering performance. Also, the posterior approximation by the employed variational Bayesian (VB) method can further decrease the estimation precision. Here, we introduce an innovative RKF method, the RKFNet, which combines the heavy-tailed-distribution-based RKF framework with the deep learning (DL) technique and eliminates the need for the precise parameters of the heavy-tailed distributions. To reduce the VB approximation error, the mixing-parameter-based function and the scale matrix are estimated by the incorporated neural network structures. Also, the stable training process is achieved by our proposed unsupervised scheduled sampling (USS) method, where a loss function based on the Student's t (ST) distribution is utilised to overcome the disturbance of the noise outliers and the filtering results of the traditional RKFs are employed as reference sequences. Furthermore, the RKFNet is evaluated against various RKFs and recurrent neural networks (RNNs) under three kinds of heavy-tailed measurement noises, and the simulation results showcase its efficacy in terms of estimation accuracy and efficiency.
\end{abstract} 

\begin{IEEEkeywords}
robust Kalman filter, heavy-tailed noise, deep learning.
\end{IEEEkeywords}

\IEEEpeerreviewmaketitle

\section{Introduction}


\IEEEPARstart{K}{alman} filtering is an algorithm used for estimating the state of a dynamic system from a series of noisy measurements over time~\cite{KF1}.
It takes into account the uncertainties associated with both the measurements and the dynamic model, and recursively updates an estimate of the state, minimising the mean squared error (MSE). Due to the versatility and effectiveness of the Kalman filter (KF), it has been a valuable tool in various fields, including target tracking~\cite{KF2}, the interest rates and inflation prediction in economics~\cite{KF3,KF4}, the GPS navigation system~\cite{KF5}, etc. While the KF yields optimal estimates under linear Gaussian models, its filtering performance may be undermined by heavy-tailed noise, which deviates from the Gaussian assumption.

Various robust Kalman filters (RKFs) have been developed to address scenarios involving heavy-tailed noise. For the M-estimator-based and outlier-detection-based RKFs, several M-estimators and outlier detectors are leveraged within the KF framework to improve the robustness of state estimation in the presence of outliers. For example, one introduces the RKFs based on the Huber function~\cite{HKF1,HKF2}, correntropy~\cite{MCKF1,MCKF2} and statistical similarity measure~\cite{SSMKF2}. Also, in~\cite{RKF-outlier-Chisquare1} and~\cite{RKF-outlier-VB2}, the Chi-Square test and variational Bayesian (VB) strategies are utilised to detect outliers and formulate novel RKF frameworks. However, their filtering performance is constrained as the stochastic characteristics of the noise are not leveraged. By contrast, the heavy-tailed-distribution-based RKFs offer a solution for the exploitation of the noise stochastic nature and provide more accurate estimation. For instance, in~\cite{RSTKF1}, the robust Student’s t-distribution-based Kalman filter (RSTKF) writes the prediction and likelihood PDFs in hierarchical Gaussian forms, and the joint posterior distribution is approximated by the VB approach~\cite{VB1}. Also, considering the skewed noise, the RKF based on the Gaussian scale mixture (GSM)  distribution (RKF-GSM) is proposed in~\cite{RKF-GSM1}, where the one-step prediction and likelihood PDFs are formulated as various skewed GSM distributions. Nonetheless, the employed VB method updates all variational parameters concurrently within the same iteration, potentially introducing instability to the estimation process. Instead, employing a heuristic approach,~\cite{RKF-GSM2} presents an elliptically-contoured-distribution-based RKF framework, where however the uncertainty of the scale matrices is not considered. Besides, in response to filtering challenges presented by the stable noise, the RKF based on sub-Gaussian $\alpha$-stable distribution~\cite{SGS1} (RKF-SG$\alpha$S) is developed in~\cite{RKF-GSM3}, and the efficient estimators of the mixing parameter are discussed. With precise noise models, the heavy-tailed-distribution-based RKFs provide robust filtering results. However, large model errors can degrade the performance of these model-based (MB) algorithms significantly.

Benefiting from the deep learning (DL) techniques~\cite{DL1}, one presents various data-driven (DD) filters, where accurate knowledge of the state-space model is not required. A prevalent method is to utilize the recurrent neural networks (RNNs), including the vanilla RNN~\cite{RNN1}, long short-term memory (LSTM) networks~\cite{LSTM1,LSTM2}, Gated recurrent units (GRUs)~\cite{GRU1}, attention mechanisms~\cite{AT1} and so on, to process the observations and generate state estimates sequentially. For example,~\cite{DDF1},~\cite{DDF2} and~\cite{DDF3} employ the LSTM, GRU and transformer for filtering or prediction tasks, respectively. However, to cope with the complexity of uncertain real-world data and achieve better belief approximation, a long latent vector is needed, which thus increases the number of network parameters and the amount of training data. By contrast, the hybrid frameworks combining RNNs with traditional filters can alleviate this drawback. For instance, the particle filter recurrent neural networks (PF-RNNs)~\cite{DDF4} combine the strengths of the RNNs and the particle filter (PF), approximating the latent state distribution as a set of particles. Without lengthening the latent vector, the required data amount is reduced. In comparison, the hybrid LSTM-KF~\cite{DDF5} learns the parameters of the KF by the LSTM and outperforms both the standalone KF and LSTM. Besides, utilising the partially known dynamics,~\cite{RKFnet1} proposes a DD-MB neural network, KalmanNet, where the structural state-space model with a dedicated RNN module is embedded in the flow of the extended KF (EKF) updating framework. Compared with traditional RNNs, the KalmanNet is more interpretable and can be trained with a smaller dataset. Also, due to the embedded RNN structure, the KalmanNet can accurately characterise the state dynamics and obtain more precise estimates than the MB filters. Nevertheless, due to the Gaussian assumption of the incorporated EKF method, the performance of the KalmanNet is limited under the heavy-tailed noise. 

In this work, we focus on a new DD-MB RKF framework, and the motivation of this work is based on the complementary advantages of the DD-MB KalmanNet and the RKFs.
\begin{enumerate}
\item The performance of the heavy-tailed-distribution-based RKFs relies on precise noise models, and extra estimation errors are caused by the posterior approximation of the employed VB method. By contrast, the KalmanNet has no requirement for precise models and can produce more accurate estimates than the MB RKFs. 
\item The KalmanNet is developed for light-tailed Gaussian noise scenarios, whereas the RKF framework can overcome the interruption of the outliers. 
\end{enumerate}
Then, we propose a hybrid framework, where the benefits of both the DD-MB strategy and the RKF framework can be leveraged. Specifically, the main contributions of this work consist in:
\begin{enumerate}
\item The DD-MB RKFNet is presented, combining the heavy-tailed-distribution-based RKF framework and the DL technique. Particularly, we consider a state-space model, where the heavy-tailed distribution of the measurement noise is expressed in a hierarchical Gaussian form without prior assumptions on its mixing density and scale matrix. Then the posterior state estimation is achieved by a KF update step with the aid of the incorporated neural networks. 

\item To improve the stability of the training process, we present an unsupervised scheduled sampling (USS) training method. In particular, the loss function based on the Student's t (ST) distribution is exploited, and the reference sequences are provided by the traditional RKFs. 

\item In the simulations, we evaluate the influences of many factors on the performance of the RKFNet. Also, a comparison between the proposed method and various benchmark filters is conducted under different heavy-tailed measurement noises.
\end{enumerate}

The remainder of this paper is structured as follows: We begin, in Section~\ref{sec:Theoretical preliminary}, with an introduction to the theoretical background, including the linear discrete-time state-space model and the deep neural network. Subsequently, Section~\ref{sec:RKFNet} delineates the architecture of the proposed RKFNet, and the USS technique is elucidated in Section~\ref{sec:USS}. Furthermore, the developed framework is assessed in target tracking scenarios in Section~\ref{sec: experiment}, whilst Section~\ref{sec:conclusion} wraps up this study with a summary.

\section{Theoretical preliminary} \label{sec:Theoretical preliminary}
In this section, we provide brief, essential details, on the key concepts, which serve as the foundation for the main developments presented in the following sections.

\subsection{Nomenclature}
\begin{table} [h!]
\centering
\begin{tabular}{p{2cm} p{6.25cm}}\toprule
\textit{Notations}    & \textit{Definitions}  \\ \toprule
$\mathcal{N}(\boldsymbol{\mu},\mathbf{\mathbf{\Sigma}})$, 
$\mathcal{N}(.;\boldsymbol{\mu},\mathbf{\mathbf{\Sigma}})$ 
             & Multivariate Gaussian pdf with mean vector $\boldsymbol{\mu}$ and covariance matrix $\mathbf{\mathbf{\Sigma}}$. \\\hline
$\mathbf{I}_n$      & $n \times n$ identity matrix. \\\hline
$|\cdot|$                  & Element-wise absolute value operation. \\\hline
$\mathrm{det}\left(\mathbf{X}\right)$ &  Determinant of square matrix $\mathbf{X}$. \\\hline
$(\cdot)^T, (\cdot)^{-1}$ &  Transpose and inverse operation.\\\hline
$\mathrm{sgn}(\cdot)$, $||\cdot||^2$ & Signum function and $L2$ norm. \\\hline
$\mathrm{exp(\cdot)}$, $\mathrm{log}(\cdot)$  & Exponential and logarithm function. \\\hline
$st(.;v,\sigma)$ & One-dimensional ST PDF with the dof value $v$ and scale parameter $\sigma$. \\\hline
$\frac{\partial(.)}{\partial (.)}$, $\propto$  &  Partial differential operation and proportional symbol.\\\hline
$\mathrm{max}(a,b)$  &  Maximum of $a$ and $b$. \\\hline
\end{tabular}
\label{table:notation}
\end{table}

\subsection{Linear Discrete-Time State-Space Model}
A linear discrete-time state-space model can be conceptualized as a probabilistic graph structure, comprising two distinct models. The signal model characterizes the state transition dynamics over time, whereas the measurement model elucidates the association between the states and their corresponding measurements. A typical linear discrete-time state-space model is expressed as follows:
\begin{equation} \label{eq:statespace_1}
\begin{cases} \mathbf{x}_k=\mathbf{F}_k\mathbf{x}_{k-1}+\mathbf{w}_{k-1}\\ \mathbf{z}_k=\mathbf{H}_k\mathbf{x}_k+\mathbf{v}_k
\end{cases}
\end{equation}
where $\mathbf{x}_k\in \mathbb{R}^n$ and $\mathbf{z}_k\in \mathbb{R}^m$ represent the hidden state and the measurement at time $k$, respectively. Also, $\mathbf{F}_k\in\mathbb{R}^{n\times n}$ and $\mathbf{H}_k\in \mathbb{R}^{m\times n}$ are the state transition and measurement matrices, respectively. Besides, $\mathbf{w}_{k}\in \mathbb{R}^n$ denotes the process noise, whilst the measurement noise is represented as $\mathbf{v}_k\in \mathbb{R}^m$.  In this work, we assume $\mathbf{w}_{k}$ follows a Gaussian distribution and $\mathbf{v}_{k}$ is heavy-tailed.

\subsection{The Deep Neural Network}
\begin{figure}[h!]
\centering
\includegraphics[width=\linewidth]{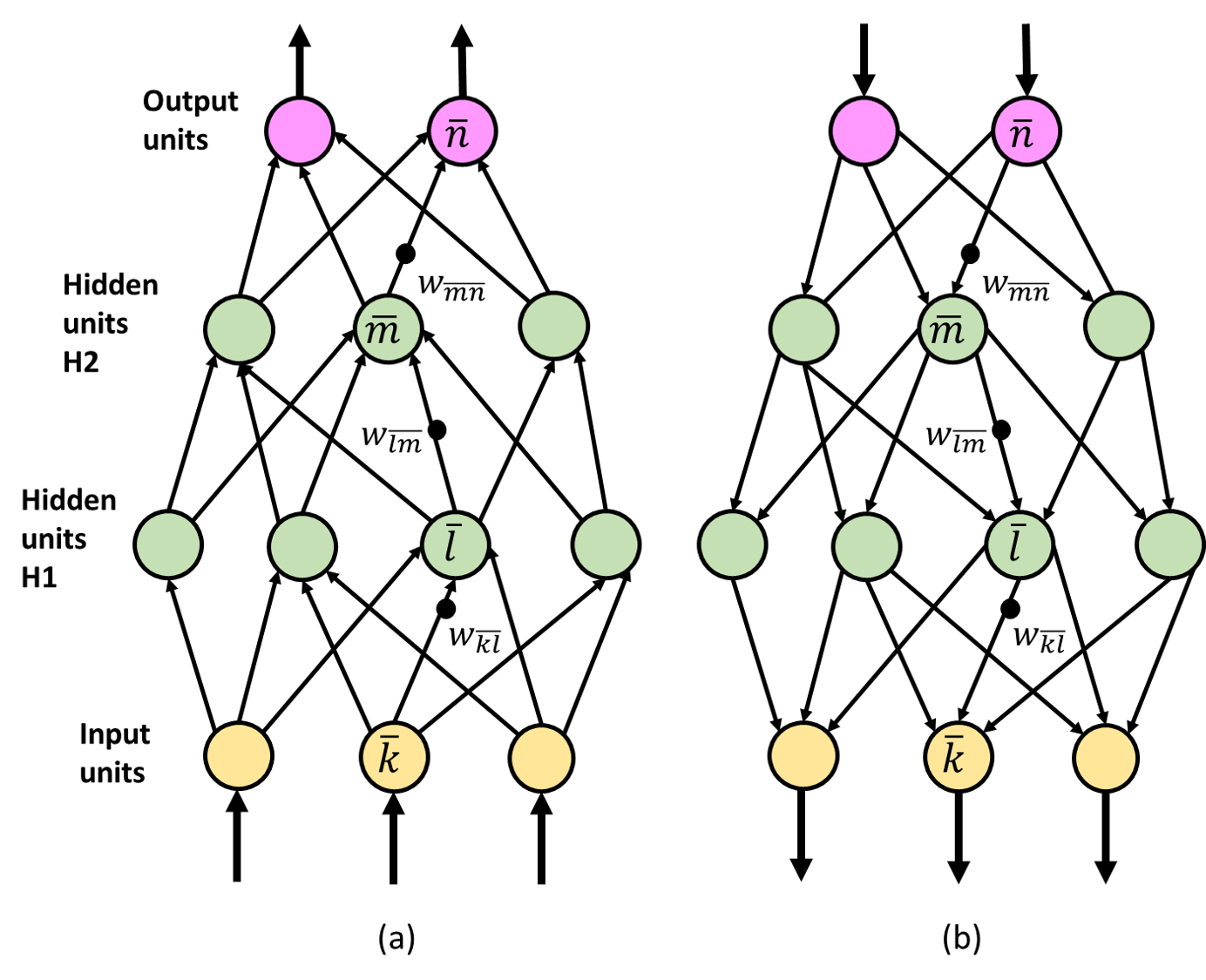}
\caption{Multilayer neural network structure. (a) and (b) plot the forward pass and backpropagation steps, respectively~\cite{DL1}.}
\label{fig:FCNN structure}
\end{figure}

Neural networks are a class of machine learning models inspired by the structure of the human brain and consist of interconnected nodes, called neurons, organized into layers. Each neuron takes inputs, performs computation and produces an output. Deep neural networks (DNNs)~\cite{DL1} are neural networks with multiple hidden layers between the input and output layers. These hidden layers enable the network to learn hierarchical representations of the input data, capturing complex patterns and relationships. Figure~\ref{fig:FCNN structure} shows an exemplary multilayer neural network, which follows a feedforward architecture. The information flows from the input layer through the hidden layers to the output layer, and each layer contains multiple neurons. For a fully connected neural network (FCNN), each neuron in one layer is connected to every neuron in the subsequent layer. 
To properly adjust the weights of neural networks, a training process based on an objective function is required, and there are two steps: forward pass and backpropagation. During the forward pass, the input data is fed into the network, and the outputs of each layer are computed sequentially as shown in Figure~\ref{fig:FCNN structure}-(a), where the neuron indexes $\overline{k}\in \mathrm{Input}$, $\overline{l}\in \mathrm{H1}$, $\overline{m}\in \mathrm{H2}$, $\overline{n}\in \mathrm{output}$. Let $(i,j) \in \{ \left(\overline{k},\overline{l}\right),\left(\overline{l},\overline{m}\right),\left(\overline{m},\overline{n}\right) \}$, we have 
\begin{equation*}
\quad z_{{j}} = \sum_{i} w_{{ij}}y_{{i}} \qquad
y_{{j}} = f\left(z_{{j}}\right). 
\end{equation*}
$y_i$ is the input of each layer and $z_j$ is the linear combination of $y_i$ based on weights $w_{ij}$. Also, $f(\cdot)$ is a nonlinear activation function with the input $z_j$ and the output $y_j$.
By contrast, Figure~\ref{fig:FCNN structure}-(b) shows the backpropagation step, where the gradients of the model parameters with respect to the loss function are calculated by the chain rule. Then, the gradient-based optimization algorithms, such as the stochastic gradient descent (SGD) and the Adam algorithm, are employed to update the parameters. The efficient backpropagation can be achieved by
\begin{equation*}
 \frac{\partial\mathcal{L}}{\partial z_{i}} = \frac{\partial\mathcal{L}}{\partial y_{i}} \frac{\partial y_{i}}{\partial z_{i}},
\end{equation*}
where 
\begin{equation*}
 \frac{\partial\mathcal{L}}{\partial y_{i}} = \sum_j w_{ij} \frac{\partial\mathcal{L}}{\partial z_{j}},
\end{equation*}
and $\mathcal{L}$ is the loss function.



\section{RKFNet architecture} \label{sec:RKFNet}
This section explains our proposed RKFNet structure. In Section~\ref{sec:new model}, we describe a state-space model, which is based on a linear model with Gaussian signal noise and unknown heavy-tailed measurement noise. Also, the proposed filtering framework is detailed in Section~\ref{sec:RKFNet filtering}. 

\subsection{A Hierarchical Gaussian State-space Model Based on The Unknown Heavy-tailed Distribution} \label{sec:new model}

We assume the zero-mean process noise follows the Gaussian distribution and the symmetric heavy-tailed distribution of the measurement noise can be written in a hierarchical Gaussian form, i.e.,
\begin{align}\label{eq:noise model}
p(\mathbf{w}_{k-1}) & = \mathcal{N}(\mathbf{0},\mathbf{Q}_{k-1}) \\
p(\mathbf{v}_{k};\mathbf{R}) &= \int_0^{+\infty} \mathcal{N}(\mathbf{v}_k;\mathbf{0},\lambda_k\mathbf{R})\pi(\lambda_k) d\lambda_k, 
\end{align}
where $\mathbf{Q}_{k-1} \in \mathbb{R}^{n\times n}$ is the covariance matrix of the state noise at time $k-1$. Also, $\mathbf{R} \in \mathbb{R}^{m\times m}$ is an unknown scale matrix, and $\lambda_k$ is a scalar mixing parameter and follows the unknown mixing density $\pi(\lambda_k)$. Then the prediction and likelihood PDF can be expressed as
\begin{align}
 p(\mathbf{x}_k|\mathbf{z}_{1:k-1})&=\mathcal{N}(\mathbf{x}_k; \mathbf{F}_k\mathbf{\hat{x}}_{k-1|k-1},\mathbf{P}_{k|k-1}) \label{eq:forcast pdf} \\ 
 p(\mathbf{z}_{k}|\mathbf{x}_k;\mathbf{R}) & =\int_0^{+\infty}\mathcal{N}(\mathbf{z}_k; \mathbf{H}_k\mathbf{x}_{k},\lambda_k\mathbf{R})\pi(\lambda_k)d\lambda_k \label{eq:likelihood pdf}
\end{align}
where $\mathbf{\hat{x}}_{k-1|k-1}$ is the posterior mean vector at time $k-1$ and the prior error covariance matrix $\mathbf{P}_{k|k-1}$ can be calculated based on the posterior error covariance matrix $\mathbf{P}_{k-1|k-1}$, i.e.,
\begin{equation} \label{eq:Pk_k-1}
 \mathbf{P}_{k|k-1}=\mathbf{F}_k\mathbf{P}_{k-1|k-1}\mathbf{F}_k^T+\mathbf{Q}_{k-1}.  
\end{equation}
\begin{rem}
In the traditional RKF frameworks, various heavy-tailed distributions have been utilised to fit the measurement noise, and the corresponding likelihood PDFs can be seen as an approximation of equation~(\ref{eq:likelihood pdf}). Specifically, $\pi(\lambda_k)$ is approximated by a fully skewed mixing density, and the uncertainty about the scale matrix $\mathbf{R}$ is represented by an inverse Wishart (IW) distribution~\cite{RKF-GSM1}. Although showing robustness and efficiency in many scenarios, the performance of the RKFs degrades when the model error is large. By contrast, the model approximation is not required in our work.
\end{rem}

\subsection{The Proposed RKFNet Filtering Framework} \label{sec:RKFNet filtering}
Based on the forecast and likelihood PDFs in~(\ref{eq:forcast pdf}) and~(\ref{eq:likelihood pdf}), the joint posterior distribution can be expressed as
\begin{equation} \label{RKFNet joint pos}
\begin{split}
p(\Tilde{\boldsymbol{\Theta}}_k|\mathbf{z}_{1:k};\mathbf{R})&\propto p(\mathbf{z}_{k}|\Tilde{\boldsymbol{\Theta}}_k;\mathbf{R})p(\Tilde{\boldsymbol{\Theta}}_k|\mathbf{z}_{1:k-1})p(\mathbf{z}_{1:k-1}) \\
 &=\mathcal{N}(\mathbf{z}_k; \mathbf{H}_k\mathbf{x}_{k},\lambda_k\mathbf{R}) \\ 
 &\times \mathcal{N}(\mathbf{x}_k; \mathbf{F}_k\mathbf{\hat{x}}_{k-1|k-1},\mathbf{P}_{k|k-1}) \\ 
 &\times \pi(\lambda_k) p(\mathbf{z}_{1:k-1}). 
\end{split}
\end{equation}
where $\Tilde{\boldsymbol{\Theta}}_k= \{\mathbf{x}_k,\lambda_k\}$ and then an RKF framework can be derived as explained in Proposition~\ref{pro:KF assimilation}:
\begin{pro}(A similar proof can be seen in Appendix C of \cite{RKF-GSM1})\label{pro:KF assimilation}
Given the posterior distribution of $\lambda_k$, 
\begin{equation} \label{eq:lamda_marginal}
p(\lambda_k|\mathbf{z}_{1:k};\mathbf{R}) = \int p(\Tilde{\boldsymbol{\Theta}}|\mathbf{z}_{1:k};\mathbf{R}) d\mathbf{x}_k,
\end{equation}
the marginal posterior distribution of $\mathbf{x}_k$ can be approximated as a Gaussian distribution, i.e.,
\begin{equation*}
p(\mathbf{x}_k|\mathbf{z}_{1:k};\mathbf{R}) \approx \mathcal{N}\left(\mathbf{x}_k;\mathbf{\hat{x}}_{k|k},\mathbf{P}_{k|k}\right),
\end{equation*}
where
\begin{equation}\label{eq:RKFNet KF update}
\begin{split}
\mathbf{\hat{x}}_{k|k-1}&=\mathbf{F}_k\mathbf{\hat{x}}_{k-1|k-1} \\ 
\mathbf{K}_{k}&=\mathbf{P}_{k|k-1} \mathbf{H}_k^T\left(\mathbf{H}_k\mathbf{P}_{k|k-1}^T\mathbf{H}_k^T+\mathbf{\Tilde{R}}_k\right)^{-1}\\
\mathbf{\hat{x}}_{k|k}&=\mathbf{\hat{x}}_{k|k-1}+\mathbf{K}_{k}\left(\mathbf{z}_k-\mathbf{H}_k\mathbf{\hat{x}}_{k|k-1}\right) \\
\mathbf{P}_{k|k}&=\left(\mathbf{I}_n-\mathbf{K}_{k}\mathbf{H}_k\right)\mathbf{P}_{k|k-1}
\end{split}
\end{equation}
and $\mathbf{K}_{k}$ denotes the Kalman gain. Also, $\mathbf{\Tilde{R}}_k$ is the modified measurement noise covariance matrix and can be written as
\begin{equation} \label{eq: RKFNet modified covariance}
\mathbf{\Tilde{R}}_k=\frac{1}{\mathrm{E}\left(\lambda_k^{-1}\right)} \mathbf{R},
\end{equation}
where 
\begin{equation*}
\mathrm{E}\left(\lambda_k^{-1}\right) = \int_0^{+ \infty} \lambda_k^{-1} p(\lambda_k|\mathbf{z}_{1:k};\mathbf{R}) d\lambda_k.  
\end{equation*}
\end{pro}
\begin{rem}
Proposition~\ref{pro:KF assimilation} provides an efficient RKF framework for the posterior estimation of $\mathbf{x}_{k}$. However, in equation~(\ref{eq: RKFNet modified covariance}), the calculation of $\mathbf{\Tilde{R}}_k$ requires the unknown $\mathbf{R}$ and $\frac{1}{\mathrm{E}\left(\lambda_k^{-1}\right)}$.
\end{rem}

To estimate $\frac{1}{\mathrm{E}\left(\lambda_k^{-1}\right)}$ and $\mathbf{R}$, and produce the posterior state estimates, we present a new neural network architecture, the RKFNet, combining the RKF framework in Proposition~\ref{pro:KF assimilation} with the DL technique. As shown in Figure~\ref{fig:RKFNet structure}, the RKFNet consists of three blocks, where Block I produces the posterior state estimation based on the KF framework in equation~(\ref{eq:RKFNet KF update}). Also, $\frac{1}{\mathrm{E}\left(\lambda_k^{-1}\right)}$ and $\mathbf{R}$ are estimated by Block II and Block III, respectively. The details of the RKFNet are explained below:

\begin{figure*}[h!]
\centering
\includegraphics[width=5.5in]{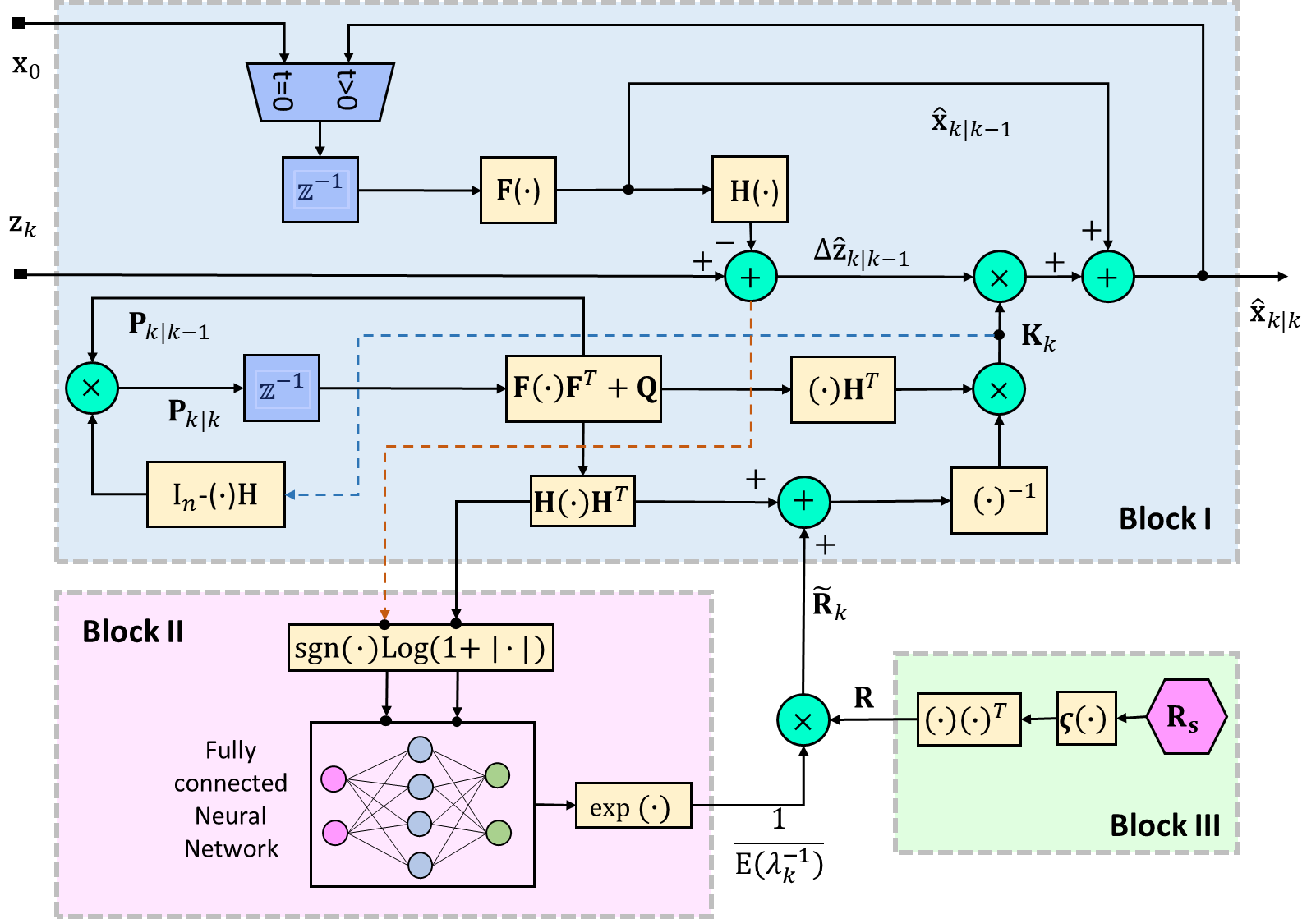}
\caption{The RKFNet structure.}
\label{fig:RKFNet structure}
\end{figure*}

1) At any time k, Block I receives the new observation $\mathbf{z}_k$ and the estimated $\mathbf{\Tilde{R}}_k$, and produces $\mathbf{\hat{x}}_{k|k}$ and $\mathbf{P}_{k|k}$.

2) To estimate $\frac{1}{\mathrm{E}\left(\lambda_k^{-1}\right)}$, we investigate the marginal posterior distribution of $\lambda_k$ in Theorem 1.
\begin{thm} \label{thm:feature selection}
At any time $k$, given the new observation $\mathbf{z}_k$, the stochastic properties of $p(\lambda_k|\mathbf{z}_{1:k};\mathbf{R})$ and $\frac{1}{\mathrm{E}\left(\lambda_k^{-1}\right)}$ are determined by $\Delta\mathbf{z}_{k|k-1}$, $\mathbf{H}\mathbf{P}_{k|k-1}\mathbf{H}^T$, $\mathbf{R}$ and $\pi(\lambda_k)$.
\end{thm} 
\textit{Proof}: See Appendix~\ref{proof:thm:marginal lamda} \qedsymbol.

According to Theorem~\ref{thm:feature selection}, Block II employs an FCNN to estimate $\frac{1}{\mathrm{E}\left(\lambda_k^{-1}\right)}$, and the input features are selected as $\Delta\mathbf{z}_{k|k-1} = \mathbf{z}_{k} - \mathbf{H}\mathbf{x}_{k|k-1}$ and $\mathbf{H}\mathbf{P}_{k|k-1}\mathbf{H}^T$. As $\mathbf{R}$ and $\pi(\lambda_k)$ are fixed, they are not incorporated as input features. Also, considering the outliers of the input features, $\mathrm{sgn}(\cdot)\mathrm{log}(1+|\cdot|)$ is employed to shrink their values. 
Moreover, to ensure a positive estimate of $\frac{1}{\mathrm{E}\left(\lambda_k^{-1}\right)}$ and narrow the range of the FCNN outputs, $\mathrm{exp(\cdot)}$ is connected to the FCNN output layer.
\begin{rem}
Due to the complexity of $p(\lambda_k|\mathbf{z}_{1:k};\mathbf{R})$, there is no close form of $\frac{1}{\mathrm{E}\left(\lambda_k^{-1}\right)}$. In the traditional RKF frameworks, an approximated PDF of $p(\lambda_k|\mathbf{z}_{1:k};\mathbf{R})$ is provided by the VB method~\cite{RKF-GSM1}, and then the estimator of $\frac{1}{\mathrm{E}\left(\lambda_k^{-1}\right)}$ can be derived. However, the VB method can introduce large approximation errors. By contrast, the RKFNet employs the neural-network-based estimator in Block II to provide better posterior estimation.
\end{rem}

3) In Block III, we introduce a neural network parameter $\mathbf{R}_s \in \mathbb{R}^{n\times n}$, and the scale covariance matrix $\mathbf{R}$ can be estimated by
\begin{equation} \label{eq:Rs}
 \mathbf{\hat{R}}=(\varsigma_1\mathbf{R}_s)(\varsigma_1\mathbf{R}_s)^T,  
\end{equation}
where $\varsigma_1 >1$ is used to increase the gradient of the loss function to $\mathbf{R}_s$. This allows for efficient updates of $\mathbf{R}_s$, particularly when the learning rate is small. Also, equation~(\ref{eq:Rs}) guarantees the symmetry and positive semi-definiteness of $\mathbf{\hat{R}}$.

\section{Unsupervised training algorithm} \label{sec:USS}
Following the introduction of the RKFNet architecture, an unsupervised training method is proposed in this section, which can benefit various practical applications in scenarios where the ground-truth data is difficult to obtain. In~\cite{RKFnet2}, the unsupervised loss function is $||\Delta\mathbf{z}_{k+1|k}||^2$. Although achieving satisfactory performance in Gaussian noise scenarios, the employed L2 norm is sensitive to the outliers of the observation prediction error $\Delta\mathbf{z}_{k+1|k}$. 

To alleviate the drawback, we adopt the function $-\mathrm{log}(st(.;v,\sigma))$ in the loss function. Given $N$ sequences with length $T$, the loss function can be written as
\begin{equation*}
\begin{split}
\mathcal{L} & \left(\overset{\smile}{\boldsymbol{\Theta}}\right) = \\
&\frac{1}{N \times T \times m}\sum_{j=1}^{N} \sum_{k=1}^T\sum_{i=1}^m -\mathrm{log}\left(st\left(\Delta\mathbf{z}_{k+1|k}^{i,j} \left(\overset{\frown}{\boldsymbol{\Theta}}\right);v,\sigma \right)\right)\\ 
&+\gamma_1 \left\| \mathrm{det} \left( \mathbf{\hat{R}} \right) -1 \right\| ^2 
+ \gamma_2 \left\| \overset{\smile}{\boldsymbol{\Theta}} \right\|^2 ,
\end{split}
\end{equation*}
where $\Delta\mathbf{z}_{k+1|k}^{i,j}$ is the $i$-th element of the observation error at time $k$ in the $j$-th trajectory. $\overset{\frown}{\boldsymbol{\Theta}}$ represents the RKFNet parameters, including the parameters of the employed FCNN and $\mathbf{R}_s$. Also, due to the difficulty of manually giving the parameters of the ST PDF, we set $v = \mathrm{exp}(\varsigma_2 v') $ and $\sigma = \mathrm{exp}(\varsigma_3 \sigma')$, where $v'$ and $\sigma'$ are specified as neural network parameters, and $\varsigma_2>1$, $\varsigma_3>1$ are introduced to increase the loss gradients with respect to $v'$ and $\sigma'$, respectively. Besides, if $\mathbf{\hat{R}}$ represents a reasonable estimate, its scaled matrix $\psi\mathbf{\hat{R}}$, where $\psi>0$, remains reasonable, indicating the existence of an infinite set of solutions. To keep the consistency of the training results, $\left\| \mathrm{det} \left( \mathbf{\hat{R}} \right) -1 \right\| ^2$ with parameter $\gamma_1$ limits the matrix determinant $\mathrm{det} \left( \mathbf{\hat{R}} \right)$ close to $1$. Furthermore, $\left\| \overset{\smile}{\boldsymbol{\Theta}} \right\|^2$ is a penalty term with parameter $\gamma_2$, and $\overset{\smile}{\boldsymbol{\Theta}}=\{\overset{\frown}{\boldsymbol{\Theta}}, v', \sigma' \}$.



\begin{figure*}[h!]
\centering
\includegraphics[width=5.5in]{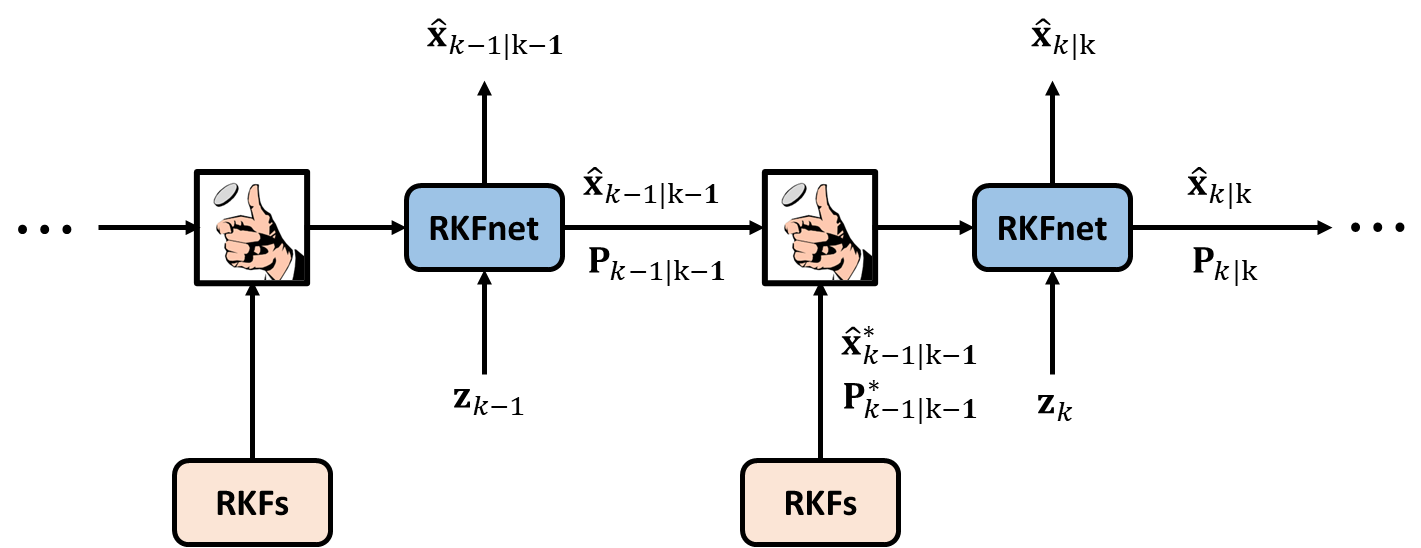}
\caption{The unsupervised scheduled sampling technique.}
\label{fig:USS}
\end{figure*}

Except for the ST-based loss function, we also present the USS method to stabilise the training process. The proposed RKFNet can be seen as a special RNN structure, where the temporally dependent information is delivered through Block I. Due to the error accumulation over time, the training process of the RKFNet is not stable especially when the noise is highly heavy-tailed. Although the scheduled sampling technique~\cite{SS1} can improve the convergence stability of the sequence-to-sequence models, it cannot be directly employed in our framework due to its requirement for ground-truth data. This drawback can be overcome by our proposed USS training method, of which the structure is shown in Figure~\ref{fig:USS}. At any time $k$, the input of the RKFNet is chosen from either $\left[\mathbf{\hat{x}}_{k-1|k-1}, \mathbf{P}_{k-1|k-1}\right]$ or the filtering result of a selected traditional RKF, $\left[\mathbf{\hat{x}}^*_{k-1|k-1}, \mathbf{P}^*_{k-1|k-1}\right]$, based on a coin toss. This increases the training stability but causes exposure bias simultaneously due to the differences between the training and inference processes. To reduce the bias, the probability of selecting the RKF estimates, $p_{t}$, decreases linearly, i.e.,
\begin{equation*}
p_{t} = \mathrm{max}(p_{min}, p_{max}-\Delta p*t)
\end{equation*}
where $t$ is the sequence iteration number and $\Delta p$ represents the decreasing rate. Also, $p_{max}$ and $p_{min}$ are the maximal and minimal probability values of $p_{t}$, respectively.

\section{NUMERICAL SIMULATIONS} \label{sec: experiment}
\subsection {The Target Tracking Model and Noises} \label{exp:model}
This section introduces the target tracking model, which is employed in the subsequent simulations and aligns with the experimental setup in~\cite{RKF-GSM3}. The state-space model is specified as equation (\ref{eq:statespace_1}), and
\begin{align}
\mathbf{F}_k=
\begin{bmatrix}
\mathbf{I}_2 & \triangle t \mathbf{I}_2\\
\mathbf{0} & \mathbf{I}_2
\end{bmatrix}, 
&\quad 
\mathbf{H}_k=
\begin{bmatrix}
\mathbf{I}_2 & \mathbf{0}
\end{bmatrix},
\end{align}
where $\triangle t=1$ is the observation interval. Also, the initial target state follows $\mathcal{N}(\mathbf{x}_{0|0}, \mathbf{P}_{0|0})$, where $\mathbf{x}_{0|0}=\left[0,0,10,10\right]^T$ and $\mathbf{P}_{0|0}=\mathrm{diag}\left(\left[25, 25, 2, 2\right]\right)$. Furthermore, the covariance matrix of the zero-mean Gaussian process noise is given by
\begin{equation} \label{eq:nominal Q} 
 \mathbf{Q}_k= 0.1*
 \begin{bmatrix}
\frac{\triangle t^3}{3}\mathbf{I}_2 & \frac{\triangle t^2}{2}\mathbf{I}_2 \\
\frac{\triangle t^2}{2}\mathbf{I}_2 & \triangle t\mathbf{I}_2
\end{bmatrix}
.
\end{equation}
Besides, three types of heavy-tailed measurement noises are chosen: Gaussian Mixture (GM), ST, and SG$\alpha$S noises. The formulation of the GM noise can be written as follows:
\begin{equation}
\mathbf{GM} = 
\begin{cases}
 \mathcal{N}(\mathbf{0},\mathbf{\overline{R}}) \quad {\rm with \ probability} \quad 0.9\\
 \mathcal{N}(\mathbf{0}, U \mathbf{\overline{R}}) \quad {\rm with \ probability} \quad 0.1,
\end{cases}
\end{equation}
where the nominal covariance matrix $\mathbf{\overline{R}}=10\mathbf{I}_2$ and the augmentation factor $U$ is selected from $[5,$ $ 10,$ $ 10^2,$ $ 10^3,$ $ 10^4,$ $ 10^5,$ $ 10^6,$ $ 10^7,$ $ 10^8]$. By contrast, both the SG$\alpha$S and ST noises utilize the scale matrix $\mathbf{\overline{R}}$, and their shape parameters are chosen from [0.3, 0.5, 0.7, 0.9, 1.1, 1.3, 1.5, 1.7, 1.85] and [0.3, 0.5, 0.7, 0.9, 1.2, 1.7, 2.5, 3.5, 6], respectively.

\subsection {Benchmark Filters} \label{Bench filters}
The proposed RKFNet is compared with 3 kinds of heavy-tailed-distribution-based RKFs\textemdash the robust Kalman filter based on the slash distribution (RKF-SL)~\cite{RKF-GSM3}, RSTKF, RKF-SG$\alpha$S. We also evaluate the performance of the RKF with true shape parameters and scale matrices of the heavy-tailed noise (RKF-TSS) and the standard KF with true noise covariance matrices (KFTNCM)~\cite{RKF-GSM2} for reference. These RKFs follow the same parameter set as in~\cite{RKF-GSM3}. Specifically, for the RKF-SG$\alpha$S, its GSIS-based variant (RKF-SG$\alpha$S-GSIS) is employed with the particle number 100, the maximal series number $\Xi=30$, the threshold for series convergence test $\varepsilon_1=10^{-2}$ and the number of latest items for series convergence test $\tau_1=4$. Besides, for the fixed-point iteration of all the RKFs, we set the maximum number of iterations $M=50$, the threshold for the convergence test $\varepsilon_2=10^{-2}$ and the number of the latest items for convergence test $\tau_2=4$. Additionally, the parameters of the heavy-tailed distributions employed in the RKFs, including the shape parameters and the scale matrices, are estimated by existing expectation maximisation (EM) and maximum likelihood estimation (MLE) methods (the detailed set is the same as in the section IV-C in~\cite{RKF-GSM3}).

\begin{figure}[h!]
\centering
\includegraphics[width=3in]{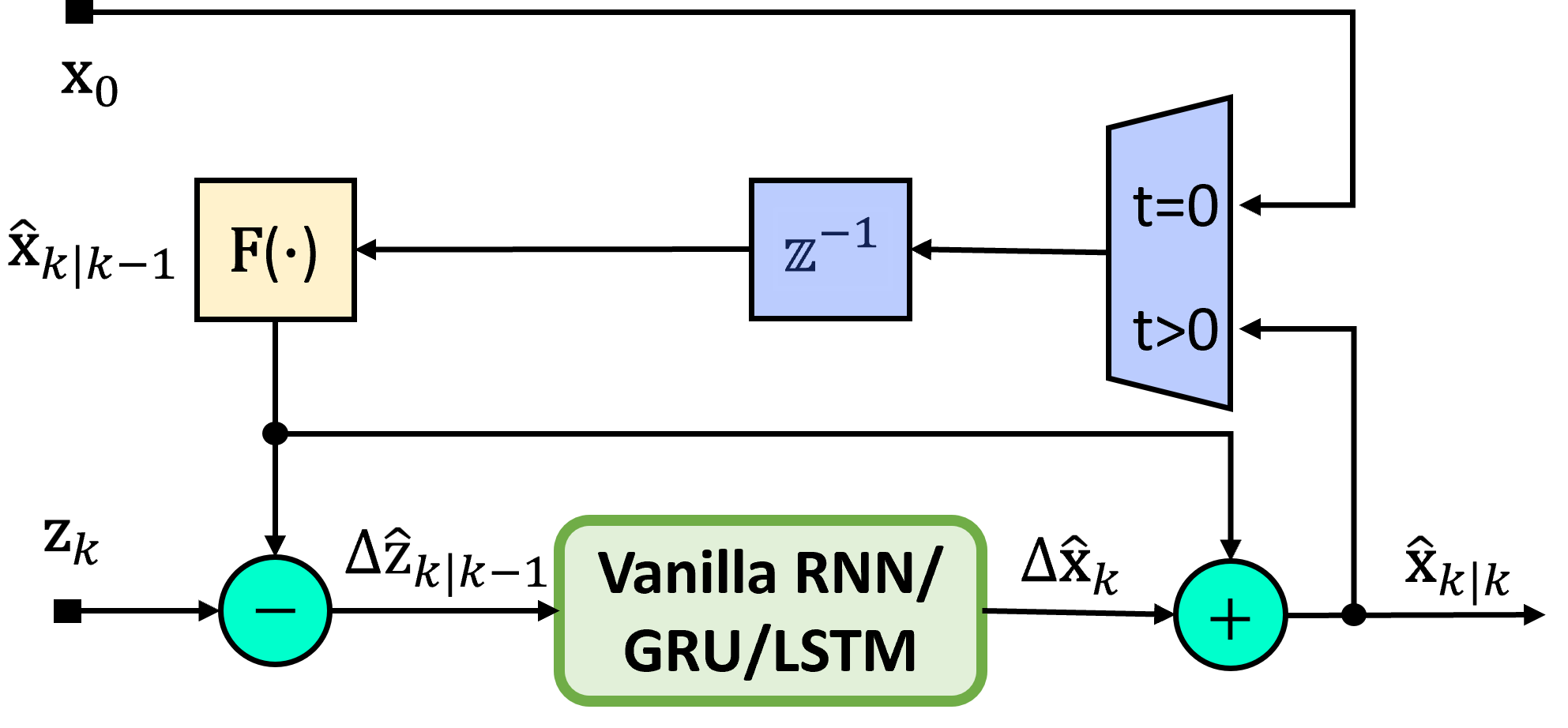}
\caption{The structure of the MB RNNs.} 
\label{fig:MB RNN}
\end{figure}

Except for the traditional RKFs, the benchmark filters also include three model-based (MB) RNNs~\cite{RKFnet1}, which employ one-layer vanilla RNN, GRU and LSTM units with hidden size 63, 36 and 31, respectively. Hence, their parameter numbers are roughly equal. As shown in Figure~\ref{fig:MB RNN}, the MB RNNs imitate the KF operation by first recovering $\mathbf{\hat{x}}_{k|k-1}$ using domain knowledge, i.e., via~(\ref{eq:RKFNet KF update}). Also, $\Delta \mathbf{\hat{z}}_{k|k-1}$ is selected as the input feature, and an increment $\Delta \mathbf{\hat{x}}_{k} = \mathbf{\hat{x}}_{k|k}-\mathbf{\hat{x}}_{k|k-1}$ is estimated by an RNN unit. 

For the RKFNet, Block II employs a 3-layer FCNN with hidden-layer sizes of 32, 64, and 32, resulting in the parameter number approximately equal to that of the MB RNNs. Also, the LeakyReLU function, characterized by a negative slope parameter of 0.1, is applied across all hidden layers. Furthermore, the elements of $\mathbf{R}_s$ are initially drawn from a uniform distribution $U[0,\frac{1}{\varsigma_1}]$, and $\varsigma_1 =300$. Besides, for the loss function, $\varsigma_2 =\varsigma_3 = 300$, $\gamma_1 = 0.1$ and $\gamma_2 = 0.0001$.

\subsection{Training Dataset and Optimiser}
In each tracking scenario, the datasets are composed of 3200 trajectories for training, 200 for cross-validation, and 200 for testing, with a fixed length of $T=100$. Also, there are 2000 iterations in every training process, and we adopt the Adam optimiser, where the learning rate starts from 0.0002 and is halved every 400 iterations. Besides, for the USS, the training batch size is 200, $p_{max}=1$ and $\Delta p = \frac{1}{600}$.

\subsection{Performance Evaluation of RKFNet across Various Factors}
In this experiment, we investigate the RKFNet filtering performance across 3 factors: $p_{min}$, the ST-based loss function and the reference trajectories produced by the traditional RKFs.

\begin{figure}[h!]
\centering
\includegraphics[width=\linewidth]{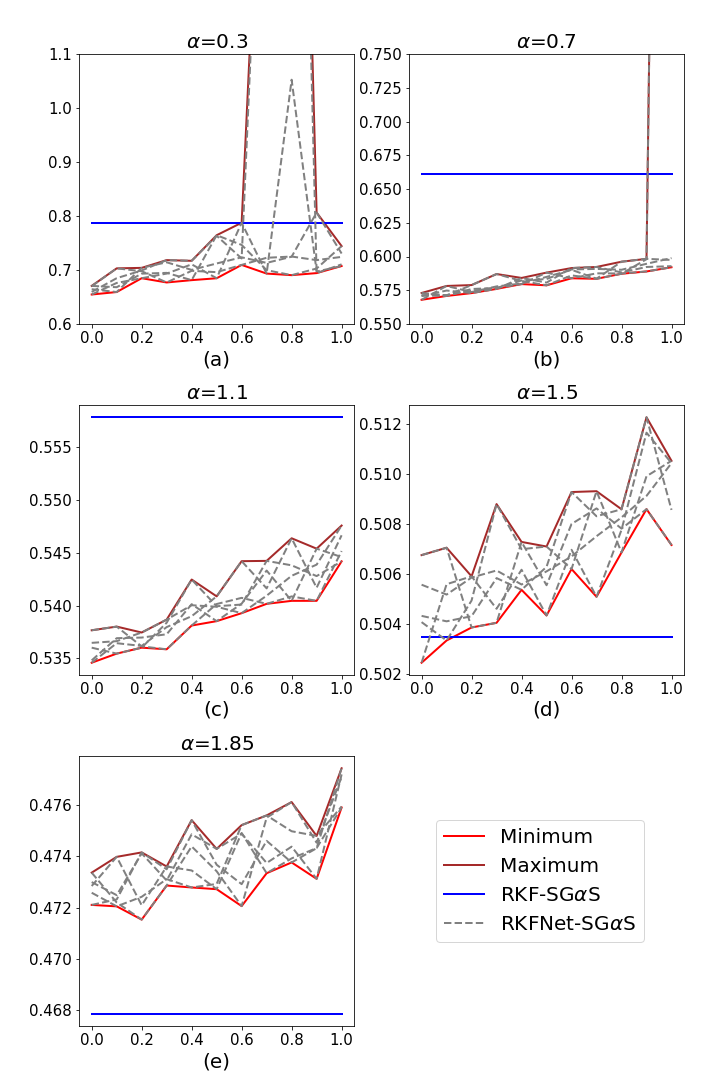}
\caption{The tracking results of the RKFNet under the SG$\alpha$S measurement noise. In every subplot, the title is the shape parameter $\alpha$ of the SG$\alpha$S distribution, the x-axis represents $p_{min}$, and the y-axis is the position estimation log10(ARMSE(m)). Also, the "RKFNet-SG$\alpha$S" represents the $5$ independent Monte Carlo results of the RKFNet, and the corresponding maximal and minimal errors are depicted by "Maximum" and "Minimum", respectively. Besides, the filtering performance of the RKF-SG$\alpha$S-GSIS is shown as a reference.}
\label{fig:p-min}
\end{figure}

1) Within the context of the target-tracking model, we evaluate the impact of $p_{min}$, ranging from 0.0 to 1.0 in the fixed increment of 0.1. Also, the SG$\alpha$S measurement noise is considered, and the shape parameter is selected from [0.3, 0.7, 1.1, 1.5, 1.85]. Besides, the filtering results of the RKF-SG$\alpha$S-GSIS are employed for the USS training method. To analyse the stability and consistency of the training results, 5 Monte Carlo experiments are run. As shown in Figure~\ref{fig:p-min}, due to the exposure errors, large $p_{min}$ can cause inaccurate and unstable filtering results, especially in the highly heavy-tailed noise scenario ($\alpha = 0.3$). By contrast, for small $p_{min}$, the RKFNet can produce more precise and reliable estimation. 

In the following simulations, $p_{min}$ is set as 0, and the RKFNet is always trained for 5 times. Also, The model corresponding to the best cross-validation results is selected as the final RKFNet model, which is then evaluated using the testing dataset. 

\begin{figure}[h!]
\centering
\includegraphics[width=\linewidth]{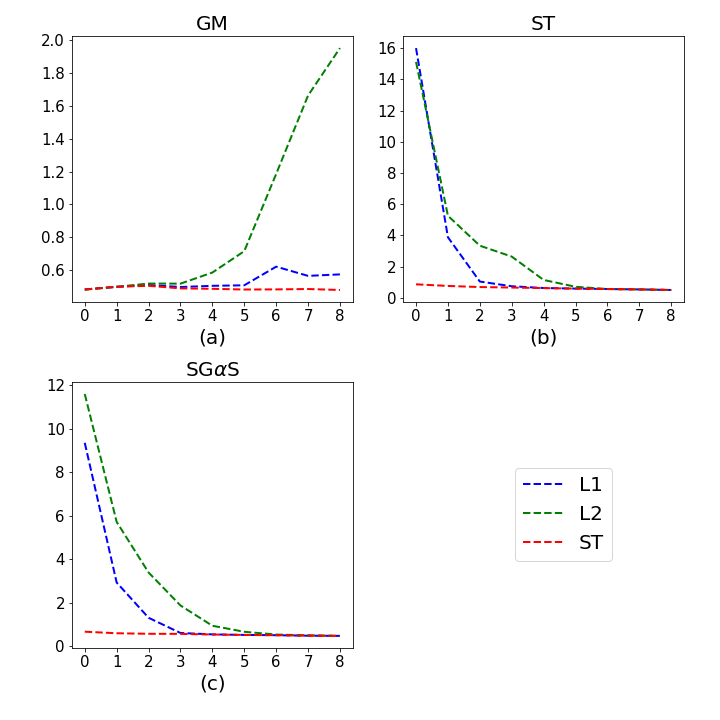}
\caption{The filtering results of the RKFNet with different loss functions. The titles of the subplots are the distributions of the heavy-tailed measurement noise. The x-axis is the shape parameter index, and the y-axis represents the position estimation log10(RMSE(m)). (a) plots the estimation errors under the GM noise, while the results under the ST and SG$\alpha$S noise are shown in (b) and (c), respectively.}
\label{fig:Loss comparision_1}
\end{figure}

\begin{figure}[h!]
\centering
\includegraphics[width=\linewidth]{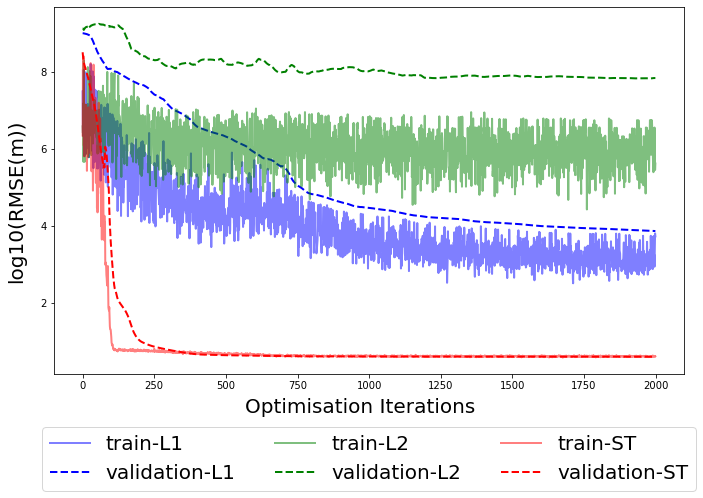}
\caption{The learning curves of the RKFNet with different loss functions under SG$\alpha$S noise $(\alpha=0.5)$. The x-axis is the optimisation iterations, and the y-axis represents the position estimation log10(ARMSE(m)).}
\label{fig:Loss comparision_2}
\end{figure}

2) The ST-based loss function is compared with the classical L1 and L2 loss functions under the three measurement noises introduced in Section~\ref{exp:model}. For light-tailed noise, the estimation errors based on the loss functions are comparable as depicted in Figure~\ref{fig:Loss comparision_1}. However, when the noise contains heavy outliers, the ST-based loss function yields the most accurate estimation due to its less sensitivity to outliers. Also, Figure~\ref{fig:Loss comparision_2} shows that the RKFNet based on the ST loss can achieve the highest convergence rate, whereas the L1/L2 losses suffer from the overfitting problem as their corresponding training results are better than the cross-validation results.

\begin{figure}[h!]
\centering
\includegraphics[width=\linewidth]{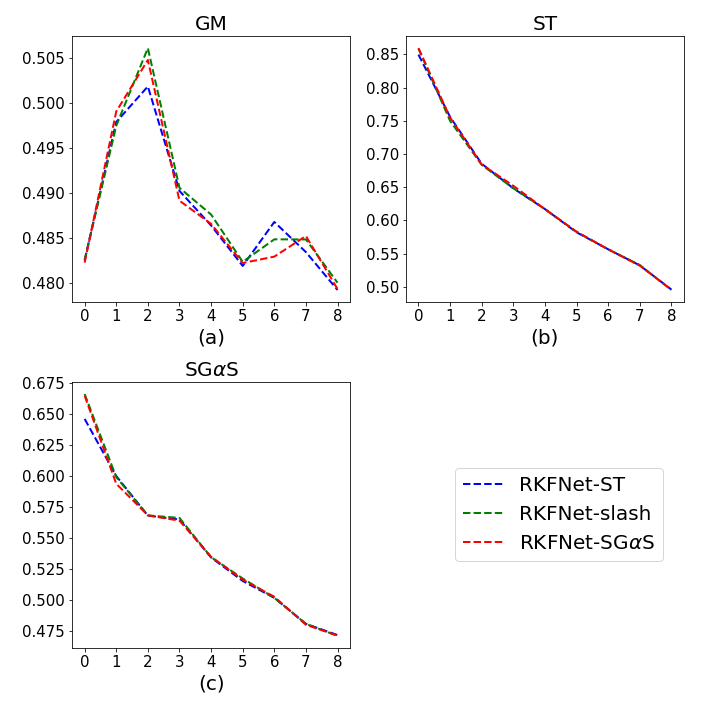}
\caption{The filtering performance comparison among the RKFNets trained based on different traditional RKFs. The titles of the subplots are the distributions of the heavy-tailed measurement noise. The x-axis is the shape parameter index, and the y-axis represents the position estimation log10(ARMSE(m)). (a) plots the estimation errors under the GM noise, while the results under the ST and SG$\alpha$S noise are shown in (b) and (c), respectively. }
\label{fig:Groundtruth comparison}
\end{figure}

3) Based on the filtering results of three RKFs, the influence of reference trajectories is assessed under the three types of heavy-tailed noises. Figure~\ref{fig:Groundtruth comparison} shows that the RKFNets trained based on different reference sequences obtain similar estimation results in all scenarios. This implies that the RKFNet has no requirement for highly precise reference sequences, which are only employed to stabilise the convergence during the initial stage of the training process. When the $p_{min}$ gradually reduces towards zero, the RKFNet can converge without the assistance of the RKF estimates.

\subsection{RKFNet vs Benchmark Filters} \label{sec:RKFNet vs RKFs}

In this experiment, our proposed RKFNet is compared with the benchmark filters introduced in section~\ref{Bench filters}. To train the RKFNet and the MB RNNs, the RKF-SG$\alpha$S is employed to produce the reference sequences. Also, we train both the RKFNet and the MB RNNs 5 times, and the final model obtains the best cross-validation results. 

\begin{figure}[h!]
\centering
\includegraphics[width=\linewidth]{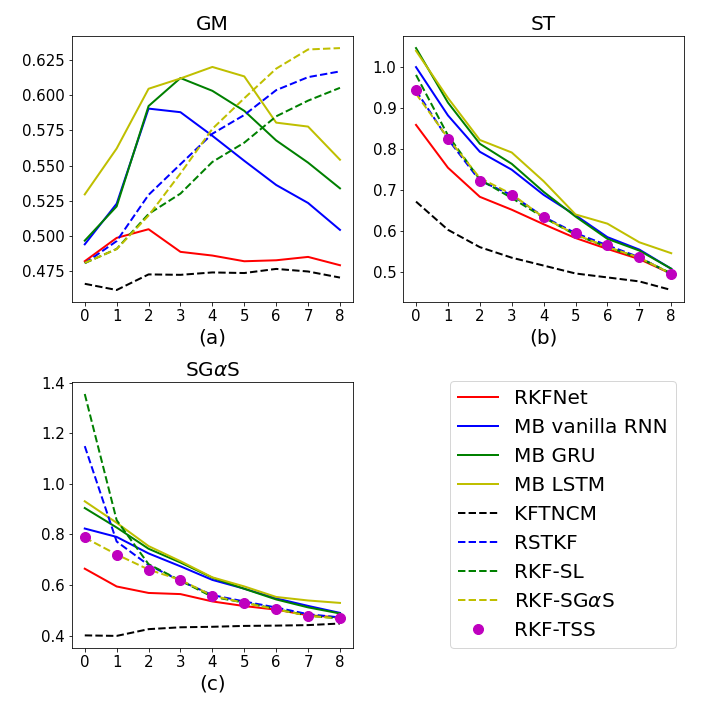}
\caption{The filtering performance comparison between the RKFNet and the benchmark filters. The titles of the subplots are the distributions of the heavy-tailed measurement noise. The x-axis is the shape parameter index, and the y-axis represents the position estimation log10(ARMSE(m)). (a) plots the estimation errors under the GM noise, while the results under the ST and SG$\alpha$S noise are shown in (b) and (c), respectively.}
\label{fig:RKFNet vs RKFs}
\end{figure}

From Figure~\ref{fig:RKFNet vs RKFs}, when the noise is light-tailed, the RKFNets and the other benchmark filters achieve similar filtering results. Nevertheless, the comparison under the heavy-tailed noise scenarios is more complicated. 1) Compared with traditional RKFs, the RKFNet can produce more precise results as it has no reliance on precise noise models. 2) The advantages of the RKFNets over the RKFs become more pronounced under the GM noise. The mixing density of the GM distribution follows a delta mixing distribution, which cannot be accurately approximated by the fully skewed mixing densities employed by the traditional RKFs. 4) Despite the true model parameters, the RKF-TSS performs worse than the RKFNet, indicating that the marginal-approximation-based VB method yields less accurate estimation than the neural-network-based method introduced in this study. 5) Compared with the MB RNNs, the RKFNet can provide better filtering results under heavy-tailed noise. Integrating the traditional RKF framework, the RKFNet diminishes the necessary neural network parameter number and achieves greater interpretability.

\begin{figure}[h!]
\centering
\includegraphics[width=\linewidth]{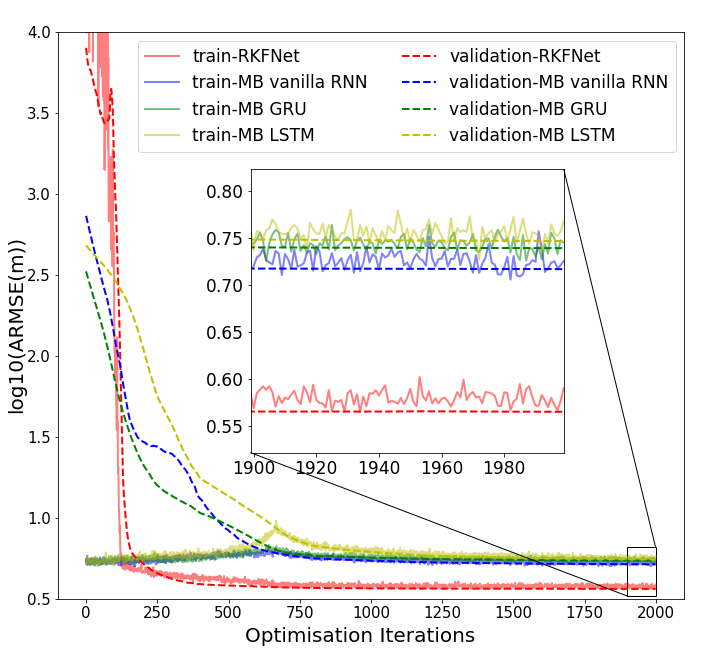}
\caption{The learning curves of the RKFNet and 3 types MB RNNs under SG$\alpha$S noise ($\alpha$=0.7). The x-axis is the optimisation iterations, and the y-axis represents the position estimation log10(ARMSE(m)).}
\label{fig:RKFNet learing curve}
\end{figure}

The training processes of the RKFNet and the MB RNNs under SG$\alpha$S noise with $\alpha$=0.7 are shown in Figure~\ref{fig:RKFNet learing curve}. The RKFNet presents a higher convergence rate and lower estimation errors than the MB RNNs. 

\begin{figure}[h!]
\centering
\includegraphics[width=\linewidth]{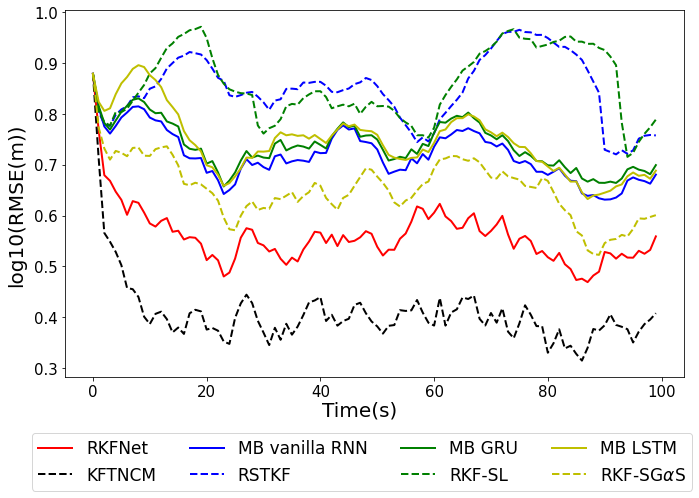}
\caption{The filtering error dynamics over time under SG$\alpha$S noise ($\alpha$=0.7). The x-axis is the tracking time, and the y-axis represents the position estimation log10(RMSE(m)).}
\label{fig:RKFNet ARMSE}
\end{figure}

Figure~\ref{fig:RKFNet ARMSE} presents the filtering error dynamics of the position estimation under the SG$\alpha$S noise with characteristic exponent $\alpha=0.7$. All the filters produce stable estimation results over time, and the RKFNet achieves the best performance.

\begin{figure}[h!]
\centering
\includegraphics[width=\linewidth]{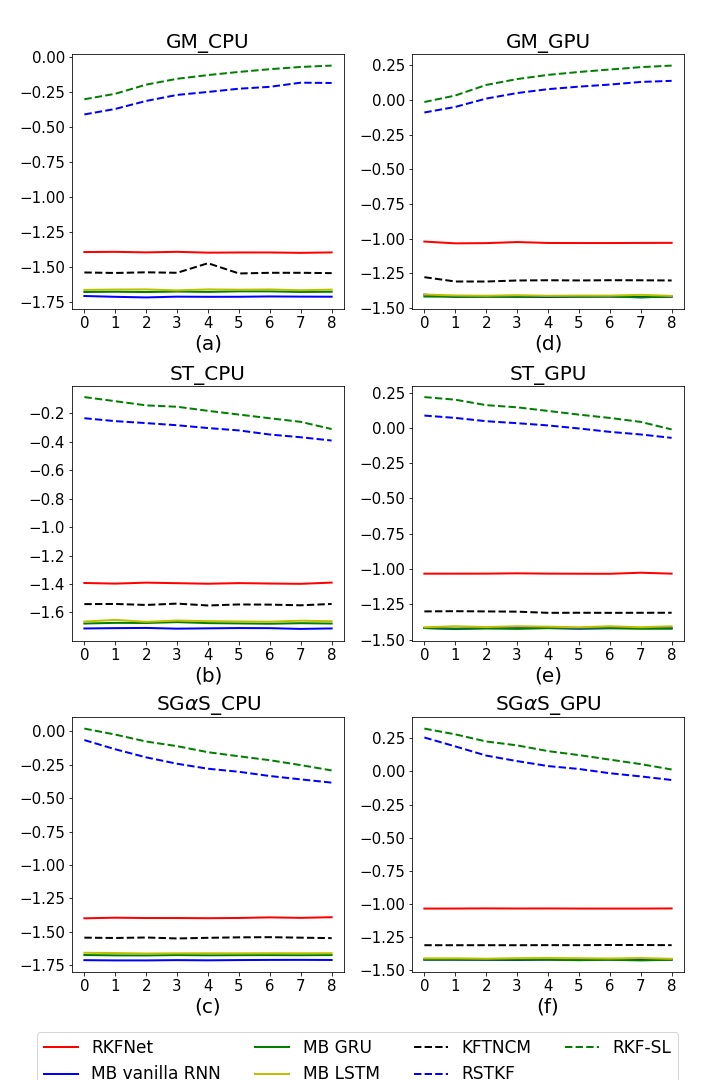}
\caption{The filtering efficiency comparison between the RKFNet and the benchmark filters. The titles of the subplots contain the distributions of the heavy-tailed measurement noise and the execution device. The x-axis is the shape parameter index, and the y-axis represents the execution time log10(time(s)). }
\label{fig:RKFNet time}
\end{figure}

Figure~\ref{fig:RKFNet time} plots the execution time of the RKFNet and the benchmark filters based on the same CPU and GPU devices under different heavy-tailed noises. Compared with RKF-SL and RSTKF, the RKFNet is more efficient as the fix-point iterations of the RKFs require more execution time. Also, as the RKF-SG$\alpha$S is more computationally expensive than the other traditional RKFs~\cite{RKF-GSM3}, its performance is not shown here. Besides, the execution time of the RKFNet is higher than MB RNNs as the KF update in Block I consists of serial operations and decreases the computational efficiency.

\section{Conclusion} \label{sec:conclusion}
In this work, we presented a DD-MB RKF framework, RKFNet, combining the traditional heavy-tailed-distribution-based RKFs with the deep-learning technique. Specifically, the mixing-parameter-based function and the scale matrix are estimated by an FCNN and an introduced neural network parameter, respectively, and then a KF update produces the posterior state estimates. Also, the USS training method is proposed to improve the stability of the training process. 

Two sets of experiments under three kinds of heavy-tailed noises are implemented to validate the proposed framework. The first simulation shows the advantages of the ST-based loss function over L1/L2 loss functions. Also, the performance of the RKFnet does not hinge on accurate reference sequences from the traditional RKF filtering results. By contrast, the second experiment set conducts a comparison between the RKFNet and the various benchmark filters. The experimental results show that in the light-tailed noise scenarios, all the filters produce similar results. However, the RKFNet achieves the best performance under heavy-tailed noises.

In the future, we will consider applying the proposed framework to more complicated scenarios. In this work, we focus on linear models with symmetric heavy-tailed measurement noise. However, an extension to linear models with skewed heavy-tailed signal and measurement noises can be considered, which can benefit more practical applications.

\appendix
\subsection{Proof of Theorem~\ref{thm:feature selection}} 
\label{proof:thm:marginal lamda} 
\begin{prf}
According to equations~(\ref{RKFNet joint pos}) and~(\ref{eq:lamda_marginal}), the marginal posterior distribution of $\lambda_k$ can be expressed as 
\begin{equation*}
\begin{split}
p(\lambda_k|\mathbf{z}_{1:k};\mathbf{R}) \propto \int & \mathcal{N}(\mathbf{z}_k; \mathbf{H}_k\mathbf{x}_{k},\lambda_k\mathbf{R}) \\
 & \times \mathcal{N}(\mathbf{x}_k; \mathbf{F}_k\mathbf{\hat{x}}_{k-1|k-1},\mathbf{P}_{k|k-1}) \\ 
 & \times \pi(\lambda_k) p(\mathbf{z}_{1:k-1}) d\mathbf{x}_k. 
\end{split}
\end{equation*}
By integrating $\mathbf{x}_k$, then 
\begin{equation*}
p(\lambda_k|\mathbf{z}_{1:k};\mathbf{R}) \propto \mathcal{N}(\Delta\mathbf{z}_{k|k-1}; \mathbf{0},\mathbf{H}\mathbf{P}_{k|k-1}\mathbf{H}^T+\lambda_k\mathbf{R}) \pi(\lambda_k),  
\end{equation*}
and hence we have Theorem~\ref{thm:feature selection}.
\qedsymbol
\end{prf}

\bibliographystyle{IEEEtran}
\bibliography{IEEEabrv,Bibliography}

\end{document}